\documentclass{emulateapj}

\shorttitle{FSRQ Gamma-ray Distributions}
\shortauthors{Singal et al.}

\slugcomment{Accepted to ApJ}

\begin{document}

\title{GAMMA-RAY LUMINOSITY AND PHOTON INDEX EVOLUTION OF FSRQ BLAZARS AND CONTRIBUTION TO THE GAMMA-RAY BACKGROUND}

\author{J. Singal\altaffilmark{1}, A. Ko\altaffilmark{2}, V. Petrosian\altaffilmark{3}}
\altaffiltext{1}{Physics Department, University of Richmond\\28 Westhampton Way, Richmond, VA 23173}
\altaffiltext{2}{Department of Physics, Masacusetts Institute of Technology \\ 77 Massachusetts Avenue, Cambridge, MA 02139-4307}
\altaffiltext{3}{Kavli Institute for Particle Astrophysics and Cosmology\\Departments of Physics and Applied Physics\\Stanford University\\382 Via Pueblo Mall, Stanford, CA 94305-4060}

\email{jsingal@richmond.edu}

\begin{abstract}
We present the redshift evolutions and distributions of the gamma-ray luminosity and photon spectral index of flat spectrum radio quasar (FSRQ) type blazars, using non-parametric methods to obtain the evolutions and distributions directly from the data.  The sample we use for analysis consists of almost all FSRQs observed with a greater than approximately 7$\sigma$ detection threshold in the first year catalog of the {\it Fermi} Gamma-ray Space Telescope's Large Area Telescope, with redshfits as determined from optical spectroscopy by Shaw et al.  We find that FSQRs undergo rapid gamma-ray luminosity evolution, but negligible photon index evolution, with redshift.  With these evolutions accounted for we determine the density evolution and luminosity function of FSRQs, and calculate their total contribution to the extragalactic gamma-ray background radiation, resolved and unresolved, which is found to be 16(+10/-4)\%, in agreement with previous studies.

\end{abstract}


\section{Introduction} \label{intro}

The majority of the extragalactic sources observed by the Large Area Telescope (LAT) on the {\it Fermi} Gamma-ray Space Telescope are blazars \citep[e.g.][]{Fermiyr1}, the type of active galactic nuclei (AGNs) in which one of the jets is aligned with our line of sight \citep[e.g.][]{BK79}.  Among AGN only blazars frequently feature prominent gamma-ray emission, and the gamma-ray emission is an essential observational tool for contstraining the physics of the central engines of AGNs \citep[e.g.][]{Dermer07}.   Understanding the characteristics of blazars is also crucial for evaluating their contribution as a source class to the extragalactic gamma-ray background (EGB) radiation.

In \citet{BP1} --- hereafter BP1 --- we explored the source counts (the so-called Log$N-$Log$S$ relation) of {\it Fermi}-LAT blazars, using those detected with a greater than approximately seven sigma detection threshold in the first-year {\it Fermi}-LAT catalog, and determined their contribution to the EGB.  Estimating the total contribution of blazars to the EGB required an extrapolation of the source counts to lower fluxes, below those detected by the {\it Fermi}-LAT.  However, in the presence of luminosity and/or density evolution with redshift, a more accurate estimate of the integrated flux from blazars requires  determining and factoring in the evolution of blazars with redshift.  

\citet{Shaw} provide spectroscopically-determined redshifts for almost all of the FSRQ blazars from the {\it Fermi}-LAT first year catalog.  With the inclusion of these redshifts, we can apply our techniques to determine the evolutions of the luminosity and photon index with redshift, the density evolution, and the distributions of luminosity and photon index for FSRQs.

Fluxes for {\it Fermi}-LAT sources are measured and reported for a given photon energy range.  The lower limit flux for detection of blazars by the {\it Fermi}-LAT depends strongly on a source's gamma-ray spectrum, such that objects with harder spectra are able to be detected above the background level at lower fluxes than those with softer spectra \citep{Atwood09}.  This means that for determination of the luminosity distribution one needs both a measure of the flux and the photon index $\Gamma$, and that then one deals with a bi-variate distribution of fluxes and indexes, which is truncated because of the this observational bias in the flux-index plane (as seen in Figure \ref{lumsandsis}), often referred to as Malmquist bias.  Additionally, of course, there is a truncation in the luminosity-redshift plane arising because of the relationship between flux and luminosity.  Obtaining a bias free determination of the distributions of luminosity and photon index is therefore necessarily quite a bit more complicated than a simple counting of sources.

\citet{Marco} use simulated data to account for the detection biases in analyzing this data.  Here we use non-parametric methods to determine the luminosity and density distributions directly from the observational data.  When dealing with a multivariate distribution, the first required step is the determination of the correlation (or statistical dependence) between the variables, which cannot be done by simple procedures when the data is truncated \citep[e.g.][]{P92}.  We use the procedures developed by Efron and Petrosian \citep[EP,][]{EP92,EP99} and extended by \citet{QP1,BP1,QP2} to account reliably for the complex observational selection biases to determine first the intrinsic correlations (if any) between the variables. These techniques have  been proven useful for application to many sources with varied characteristics, including to the Log$N-$Log$S$ relation for blazars in BP1, and to radio and optical luminosity in quasars in \citet{QP1} and \citet{QP2}, where references to earlier works are presented.

In this paper we apply these methods to determine the luminosity and photon index evolutions of {\it Fermi}-LAT blazars, as well as the density evolution, and local ($z$=0) gamma-ray luminosity function.   In \S \ref{datasec} we discuss the data used, and in \S \ref{evs} we explain the techniques used and present the results for the luminosity and photon index evolution.  In \S \ref{dev} and \S \ref{ll} we present the density evolution and local luminosity function and photon index distribution.  In \S \ref{bgnd} we calculate the total contribution of FSRQs to the extragalactic gamma-ray background (EGB) radiation. This paper assumes the standard cosmology throughout.

\section{Data}\label{datasec}

For this analysis we use the FSRQ blazars reported in the {\it Fermi}-LAT first year extragalactic source catalog \citep[e.g][]{FermiAGN} that have a detection test statistic $TS \geq 50$ and which lie at Galactic latitude $\vert b \vert \geq 20^{\circ}$.   The test statistic is defined as $TS = -2 \, \times \, (\ln (L_0) - \ln (L_1))$, where $L_0$ is the likelihood of the source being not actually present and the flux being due only to the background in that location (null hypothesis) and $L_1$ is the likelihood of the the hypothesis being tested, which is that the source is present along with the background in that location.  The significance level of a given detection is approximately $n \times \sigma = \sqrt{TS}$ .  Of 425 total of such extragalactic sources, 325 are identified as blazars.  Spectroscopically determined redshifts are provided for 184 of the FSRQ-type blazars in \citet{Shaw}, which is all or nearly all of the FSRQ-type blazars in the $TS \geq 50$ and $\vert b \vert \geq 20^{\circ}$ sample, making the sample complete for spectroscopic redshifts.\footnote{While the {\it Fermi}-LAT first year extragalactic source catalog identified 161 blazars in the $TS \geq 50$ and $\vert b \vert \geq 20^{\circ}$ sample as FSRQs, \citet{Shaw} spectroscopically classify 23 additional blazars in this sample as FSRQs which were identified in the {\it Fermi}-LAT catalog as being of unknown type.}

\begin{figure}
\includegraphics[width=3.5in]{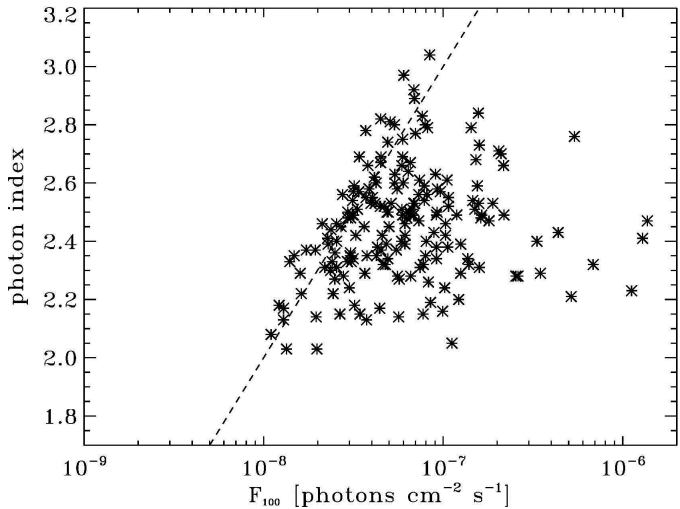}
\caption{ Flux and photon spectral index for the 184 {\it Fermi}-LAT first year FSRQ blazars used in this analysis, those with $TS$ $\geq$ 50 and $\vert b \vert \geq 20^\circ$ and known spectroscopic redshifts provided by \citet{Shaw}.   It is seen that there is a selection bias against soft spectrum sources at fluxes below $\sim 10^{-7}$ photons cm$^{-2}$ sec$^{-1}$.  We also show the line used for the truncation boundary in this analysis, as discussed in \S \ref{datasec} and determined in BP1. }
\label{lumsandsis}
\end{figure}

The 184 FSRQ blazars used here range in gamma-ray flux in the range from 100 MeV to 100 GeV from $1.1 \times 10^{-8}$ to $1.37 \times 10^{-6}$ photons cm$^{-2}$ sec$^{-1}$, and the flux in this range is designated as $F_{100}$.  The photon index $\Gamma$ is defined such the number of photons as a function of photon energy is given by $n\!(E)dE \propto E^{-\Gamma}$ (or the $\nu F_{\nu} \propto \nu^{-\Gamma +2}$), and is obtained by fitting a simple power-law to the spectra in the above energy interval. The photon index is reported directly in the {\it Fermi-LAT} source catalog and in the present sample ranges from 2.03 to 3.04.  We recover $F_{100}$ from the reported flux density ($K$), pivot energy ($E_p$), and photon index with

\begin{equation}
F_{100} = \int_{100 MeV}^{100 GeV} \, K \, {\left( E \over E_ p \right)}^\Gamma \, dE
\label{F100}
\end{equation}
The bias mentioned above, stemming from the dependence of the {\it Fermi}-LAT point spread function (PSF) with energy, is apparent, as there is a strong selection against soft spectrum sources at fluxes below $F_{100} \sim 10^{-7}$ photons cm$^{-2}$ sec$^{-1}$.  

Each source has an associated $TS$ as discussed above, and the sources' TS values vary in part because the background flux is a function of position on the sky, as discussed in \citet{FermiAGN}.  The approximate limiting flux for inclusion in the survey, then, is given by $F_{\rm lim}=F_{100} / \sqrt{TS/50}$, where $TS$ is a function of the position on the sky and the photon index $\Gamma$.  However, as discussed in BP1, because the limiting flux as determined in this way is not the optimal estimate, we use a more conservative truncation as shown by the straight line in Figure \ref{lumsandsis}. In BP1 we derive the optimal location of this truncation line for a $TS \geq 50$ sample from the {\it Fermi}-LAT first year extragalactic source catalog.  With the present FSRQ sample, applying the truncation line excludes 33 sources from the analysis.  As discussed in BP1, the optimal location for the truncation line may exclude some sources leading to increased statistical uncertainty, but provides increased accuracy of results.  

The FSRQs in the sample range in redshift from 0.001 to 3.197.  To convert between photon flux $F_{100}$ and the gamma-ray luminosity $L_{\gamma}$, one must first convert from $F_{100}$ to energy flux ($E_{100}$) with
\begin{equation}
{ {E_{100}} \over {F_{100} } }\equiv R(\Gamma) \cong 100 \times {{\Gamma-1} \over {\Gamma-2}} \times {{1-10^{3(2-\Gamma)}} \over {1-10^{3(1-\Gamma)}} } \,\,\,{\rm MeV / photon},
\label{econv}
\end{equation}

(except for $\Gamma=2$ and $\Gamma=1$ for which $R(2)=\ln 10^3/(1-10^{-3})\sim 6.9$ and  $R(1)=(10^3-1)/\ln 10^3\sim 150$ respectively) and then from energy flux to luminosity with 
\begin{equation}
L_{\gamma}=4\, \pi \, {D_L (z)}^2 \, K(z) \, E_{100}
\label{lconv}
\end{equation}
where $D_L(z)$ is the luminosity distance determined from the standard cosmological model, and $K(z)$ is the K-correction factor given by $K(z)=(1+z)^{\Gamma-2}$.  The gamma-ray luminosities of FSRQs in the sample range from 1.95$\times$10$^{40}$ erg sec$^{-1}$ to 6.32$\times$10$^{49}$ erg sec$^{-1}$.  A nearly identical sample has been analyzed by \citet{Marco} which finds agreement with the major conclusions of this work.

\section{Luminosity and Photon Index Evolutions}\label{evs}

\subsection{Luminosity and density evolution}\label{lde}

The LF expresses the number of objects per unit comoving volume $V$ per unit source luminosity, so that, if there are no other correlated parameters, the number density of objects is $dN/dV = \int dL \Psi(L, z)$ and the total number is $N = \int dL \, \int dz \, (dV/dz) \, \Psi(L, z)$.  To provide for luminosity and density evolution, without loss of generality, we can write a LF in some waveband $a$ as 

\begin{equation}
\Psi_{a}\!(L_a,z) = \rho(z)\,\psi_a\!(L_{a}/g_{a}\!(z) , \eta_a^j)/g_a\!(z),
\label{lumeq}
\end{equation}
where $g_{a}\!(z)$ describes the luminosity evolution with redshift and $\rho(z)$  describes the comoving density evolution with redshift, and $\eta_a^j$ stands for parameters that describe the shape (e.g. power law indices and break values) of the $a$ band LF.   If the parameters $\eta_a^j$ have some redshift dependence this is equivalent to having luminosity dependent density evolution, and if $\eta_a^j$ have some luminosity dependence this is equivalent to luminosity dependent luminosity evolution.  The methods used here for determining intrinsic correlations and distributions have been demonstrated through previous works and analysis of simulated data sets for the case where the parameters $\eta_a^j$ are constant in redshift, i.e. the luminosity function is modeled with only luminosity evolution and density evolution.  In principle the methods could be extended to allow for evolving parameters $\eta_a^j$ with redshift or with luminosity dependence.  Here we start with the simpler model where there is only luminosity evolution and density evolution to capture the most important features of the evolution of the luminosity function, and find that it is adequate to do so.  We consider this form of the LF for the gamma-ray luminosity, and a similar form for the photon index distribution.

\begin{figure}
\includegraphics[width=3.5in]{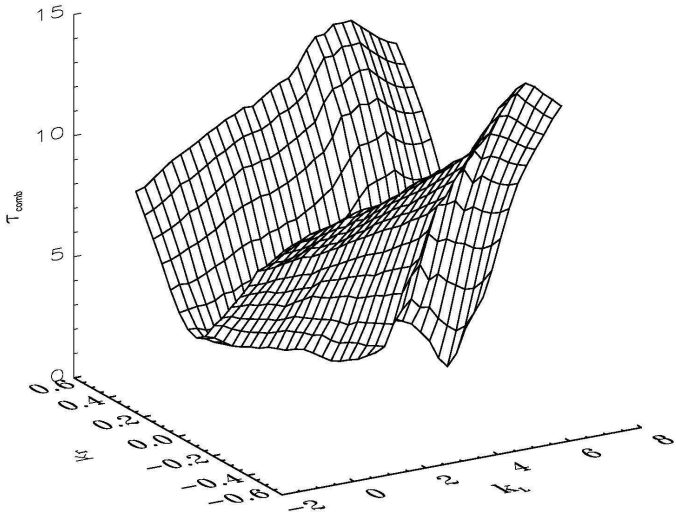}
\caption{Surface plot of the value of $\tau_{\rm comb}$ showing the location of the minimum region where the favored values of $k_{\rm L}$ and $k_{\rm \Gamma}$ lie, for the forms of the evolutions  given by equation \ref{Levolutionold}. }
\label{tauass}
\end{figure}

\begin{figure}
\includegraphics[width=3.5in]{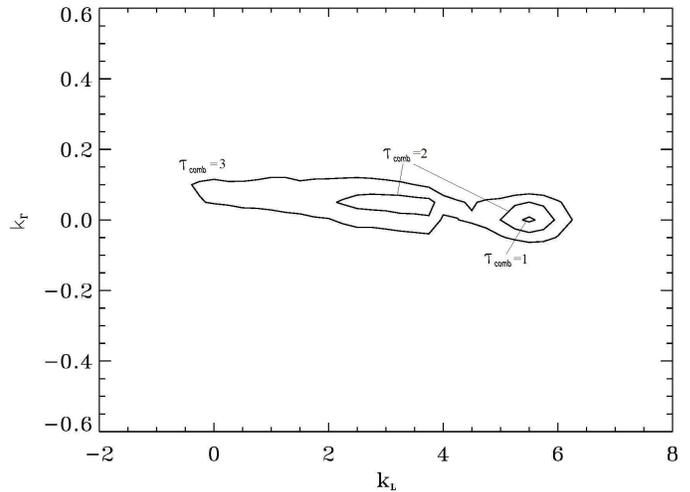}
\caption{The 1, 2, and 3 $\sigma$ contours of $\tau_{\rm comb}$ for the simultaneous best fit values of $k_{\rm L}$ and $k_{\rm \Gamma}$ for the forms of the evolutions  given by equation \ref{Levolutionold}.   }
\label{alphas}
\end{figure} 

Once the luminosity evolution $g_{a}\!(z)$ is determined using the EP method we can obtain the mono-variate distributions of the independent variables $L'_{\rm a}=L_{\rm a}/g_{\rm a}(z)$ and $z$, namely the density evolution $\rho(z)$ and local ($z$=0) LF $\psi_{\rm a}$.  The total number of observed objects seen is then
\begin{equation}
N_{tot} = \int_0^{z_{max}} dz \int_{L_{\rm min}(z)}^\infty dL_{a} \int {  d\Omega \, \, \rho(z) \, {dV \over dz} \, { {\psi_a\!\left(L_{a}/g_{a}\!(z)\right)} \over {g_a\!(z)} }  } ,
\label{inteq}
\end{equation}
In this case, because $L_{\rm min}(z)$ has some dependence on $\Gamma$, an integral over the distribution of $\Gamma$ must be performed as well, leading to
\begin{equation}
N_{tot} = \int_0^{z_{max}} dz \int_{L_{\rm min}(z, \, \Gamma)}^\infty dL_{\gamma} \int d\Gamma \int {  d\Omega \, \, \rho(z) \, {dV \over dz} \, { {\psi_{\rm \gamma}\!\left(L_{\gamma}/g_{\rm L}\!(z)\right)} \over {g_{\rm L}\!(z)} } h(\Gamma) } ,
\label{inteq2}
\end{equation}
where the distribution $h(\Gamma)$ should have a norm of 1.\footnote{We note that this form for equation \ref{inteq2} is only strictly true if $h(\Gamma)$ is  independent of redshift and luminosity.  The former is seen in \S \ref{evsec} while the later is discussed in \S \ref{bgnd}. }

Because the sample at hand has a mean redshift of around $z$=1, we can assume a simple power law for the evolutions within this redshift range

\begin{equation}
g_{\rm a}(z)=(1+z)^{k_{\rm a}}.
\label{Levolutionold}
\end{equation}
We have also considered a more complicated parameterization which allows a non-monotonic form with a turnover at a critical redshift

\begin{equation}
g_{\rm a}(z)={ { (1+z)^{k_{\rm a}} } \over { 1 + \left({  {1+z} \over {z_{cr}}  }  \right)^{k_{\rm a}} } },
\label{Levolution}
\end{equation}
however, the form of equation \ref{Levolutionold} provides an valid fit to the data as determined by the methods of \ref{evsec}, so we choose the simpler parameterization.

We discuss the determination of the evolution factors $g_a(z)$ with the EP method, which in this parameterization becomes a determination of $k_a$, in \S \ref{evsec}.  The density evolution function $\rho(z)$ is determined by the method discussed in \S \ref{dev}.  Once these are determined we construct the local (de-evolved) LF $\psi_{\rm L_{\gamma}'}\!(L_{\gamma}')$, shown in \S \ref{ll}.

\subsection{Determination of best fit correlations} \label{evsec}

Here we first give a brief summary of the algebra involved in the EP method.  This method finds the best-fit values of parameters describing correlations between variables by removing the correlation with some function and testing for independence.  We utilize a modified version of the Kendall tau test to estimate the best-fit values of parameters describing the correlation functions between variables, using the test statistic 

\begin{equation}
\tau = {{\sum_{j}{(\mathcal{R}_j-\mathcal{E}_j)}} \over {\sqrt{\sum_j{\mathcal{V}_j}}}}
\label{tauen}
\end{equation}
to quantify the independence of two variables in a dataset, say ($x_j,y_j$) for  $j=1, \dots, n$.  Here $\mathcal{R}_j$ is the dependent variable ($y$) rank of the data point $j$ in a set associated with it.  The expectation value of the rank is $\mathcal{E}_j=(1/2)(n+1)$ and the variance is $\mathcal{V}_j=(1/12)(n^{2}-1)$.  For untruncated data (i.e. data truncated parallel to the axes) the set associated with point $j$ includes all of the points with a lower  (or higher) independent variable value ($x_k < x_j$).  If the data is truncated the unbiased set is then the {\it associated set} consisting only of those points of lower (or higher) independent variable ($x$) value that would have been observed if they were at the $x$ value of point $j$ given the truncation.  A very simple example of an associated set is discussed in the Appendix.

If ($x_j,y_j$) are uncorrelated then the ranks of all of the points $\mathcal{R}_j$ in the dependent variable within their associated set should be distributed uniformly between 1 and the number of points in the set $n$, with the rank uncorrelated with their independent variable value. Then the points' contributions to $\tau$ will tend to sum to zero.   On the other hand, if the indepenent and dependent verables are correlated, then the rank of a point in the dependent variable will be correlated with its  independent variable, and because the set to be ranked against consists of points with a lower independet variable value, the contributions to $\tau$ will not sum to zero.

Independence of the variables is rejected at the $m \, \sigma$ level if $\vert \, \tau \, \vert > m$, and this can be considered the same standard deviation as would be calculated from another method such as least-squares fitting, as discussed in \citet{EP99}.   If the variables are not independent, to find the best fit correlation the $y$ data are then adjusted by defining $y'_j=y_j/F(x_j)$  and the rank test is repeated, with different values of parameters of the function $F$. In this case $F(x_j)$ would be the luminosity or photon index evolution factors, $g_{L}(z)$ and $g_{\Gamma}(z)$, with the forms specified by equation \ref{Levolutionold}.

The procedure for determining simultaneously the best fit $k_{\rm L}$ and $k_{\rm \Gamma}$ is more complicated because we now are dealing with a three dimensional distribution  ($L_{\gamma}$, $\Gamma$, $z$) and two correlation functions ($g_{\rm L}\!(z)$ and $g_{\rm\Gamma}\!(z)$).  The associated set for any object $i$ then consists of only those objects which would still be present in the survey if they were located at the redshift of object $i$ , given the luminosity and photon index evolution factors and in light of the truncation boundary in the $F_{100}$-$\Gamma$ plane.

We form a combined test statistic $\tau_{\rm comb} = \sqrt{\tau_{\rm L}^2 + \tau_{\rm \Gamma}^2} $ where $\tau_{\rm L}$ is that calculated with equation \ref{tauen} with $L_{\gamma}$ as the independent variable and $z$ as the dependent variable, and $\tau_{\rm \Gamma}$ is that calculated with $\Gamma$ as the independent variable and $z$ as the dependent variable.  Because the luminosity corresponding to a given measured flux and redshift is dependent on the photon index, adjusting the luminosity or the photon index by a redshft dependent factor (i.e. performing the $L_{\rm \gamma}'=L_{\rm \gamma}/g_{\rm L}(z)$ and $\Gamma'=\Gamma/g_L(z)$) affects the other variable.  Therefore, $\tau_{\rm L}$ and $\tau_{\rm \Gamma}$ must be evaluated simultaneously for every $L_{\gamma}'$ and $\Gamma'$ combination.  We perform the evaluations on a grid of $L_{\rm \gamma}'$ and $\Gamma'$ values, i.e. a grid of $k_{\rm L}$ and $k_{\rm \Gamma}$ values with spacing as small as 0.05.  As with the one dimensional case, the best fit values of $k_{\rm L}$ and $k_{\rm \Gamma}$ are those that minimize $\tau_{\rm comb}$.  Since $\tau_{\rm comb}$ is a geometric average and cannot be negative, the $m \, \sigma$ ranges of the best-fit values of $k_{\rm L}$ and $k_{\rm \Gamma}$ are those that lead to $\tau < m$.  

The favored values of $k_{\rm L}$ and $k_{\rm \Gamma}$ are those that simultaneously give the lowest $\tau_{\rm comb}$ and, again, we take the $1 \sigma$ limits as those in which $\tau_{\rm comb} \, \leq 1$.  For visualization, Figure \ref{tauass} shows a surface plot of $\tau_{\rm comb}$.  Figure \ref{alphas} shows the best fit values of $k_{\rm L}$ and $k_{\Gamma}$ taking the 1, 2, and 3 $\sigma$ contours.   We have verified this general method with a simulated Monte Carlo datasets as discussed in \citet{QP1} and BP1.  We see that strongly positive evolution in gamma-ray luminosity is favored, along with no evolution of the photon index.

\begin{figure}
\includegraphics[width=3.5in]{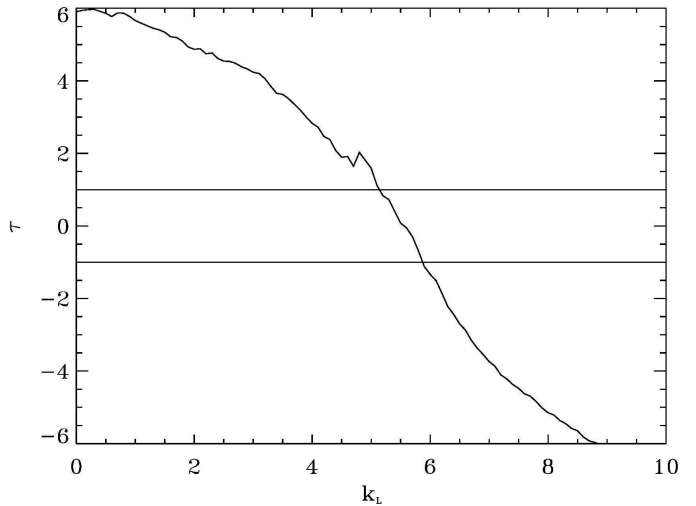}
\caption{$\tau$ versus $k_L$ for the set of objects where $F_{100} \geq 10^{-7}$ photons cm$^{-2}$ and truncation in the $F_{100}$-$\Gamma$ plane is unimportant, so determination of the luminosity evolution reduces to a simpler 2-dimensional correlation.  The best-fit 1$\sigma$ range of $k_L$ is where $|\tau| \leq 1$, and this results in $k_L$=5.5$\pm$0.5, in good agreement with the results when the full 3-dimensional case is considered with the whole data set (e.g. Figure \ref{alphas}). }
\label{simpcase}
\end{figure} 

The result for the luminosity evolution can be checked by considering only those objects with $F_{100} \geq 10^{-7}$ photons cm$^{-2}$ sec$^{-1}$, where the truncation in the $F_{100}$-$\Gamma$ plane is not relevant (see Figure \ref{lumsandsis}).  The determination of the luminosity evolution in this case reduces to a simpler 2-dimensional determination of the correlation between $L_{\gamma}$ and $z$.  The associated set for each object $i$ is then those objects which would still be present in the survey if they were located at the redshift of object $i$ given the luminosity evolution factor and a simple limiting flux of $F_{100} \geq 10^{-7}$ photons cm$^{-2}$ sec$^{-1}$.  Figure \ref{simpcase} shows $\tau$ versus $k_L$ for this subset, and the results indicate that $k_L$=5.5$\pm$0.5, in good agreement with the results when the full 3-dimensional case is considered with the whole data set.  \citet{Marco} also derive quite strong FSRQ  luminosity evolution with redshift, although with a more complicated parameterization.  We note in passing that, as can be seen from the contours in \ref{alphas}, a complete lack of luminosity evolution is not ruled out at the 3 $\sigma$ level.  However the results from considering the untruncated subset, along with those in other works, strongly favors positive luminosity evolution.

\section{Density evolution} \label{dev}

We turn now to the density evolution $\rho(z)$.  The cumulative density function is

\begin{equation}
\sigma(z) = \int_0^z { {{dV} \over {dz}} \, \rho\!(z) \, dz}
\end{equation}
which, following the procedure in \citet{P92} based on \citet{L-B71}, can be calculated by

\begin{deluxetable}{lcccccccc}\label{tizzable}
\tabletypesize{\scriptsize}
\tablecaption{Coefficients for polynomial fit\tablenotemark{a} to density evolution $\rho(z)$ vs. $z$  } 
\tablecolumns{3}
\startdata
  & $z \leq 0.75$ & $z \geq 0.75$ \\
\hline
c & +8.9x10$^{-8}$ & $-$6.0x10$^{-9}$ \\
$a_1$ & $-$4.4x10$^{-8}$   & +3.4x10$^{-7}$  \\
$a_2$ & +3.0x10$^{-7}$   & $-$3.0x10$^{-4}$  \\
$a_3$ & $-$2.5x10$^{-7}$   & +6.5x10$^{-8}$  \\
\enddata
\tablenotetext{a}{Polynomial fits are of the form $\rho(z) = c_1 \, + \, a_1 \, z \, + \, a_2 \, z^2 \, + \, a_3 \, z^3  $.}
\end{deluxetable}

\begin{equation}
\sigma(z) = \prod_{j}{\left(1 + {1 \over m(j)}\right)}
\label{sigmaeqn}
\end{equation}
In this case $j$ runs over all objects with a redshift lower than or equal to $z$, and $m(j)$ is the number of objects with a redshift lower than the redshift of object $j$ {\it which are in object j's associated set}. The associated set for object $j$ consists of those objects which would still be in the survey if they had object $j$'s luminosity.  Then the differential density evolution $\rho(z)$ is 

\begin{equation}
\rho\!(z) = {d \sigma\!(z) \over dz} \times {1 \over dV/dz}
\label{rhoeqn}
\end{equation}
If $\sigma(z)$ is expressed in number of objects less than redshift $z$ per solid angle (N$<z$ sr$^-1$), then $\rho(z)$ here is expressed in N/dz Mpc$^{-3}$.  

However, to determine the density evolution, the previously determined luminosity evolution must be factored out of each objects' luminosity.  Thus, the luminosities for determining inclusion in the associated sets for each object in the calculation of $\sigma$ by equation \ref{sigmaeqn} are scaled by taking out factors of $g_{\rm L}\!(z)$ and $g_{\rm \Gamma}\!(z)$, determined as above.

\begin{figure}
\includegraphics[width=3.5in]{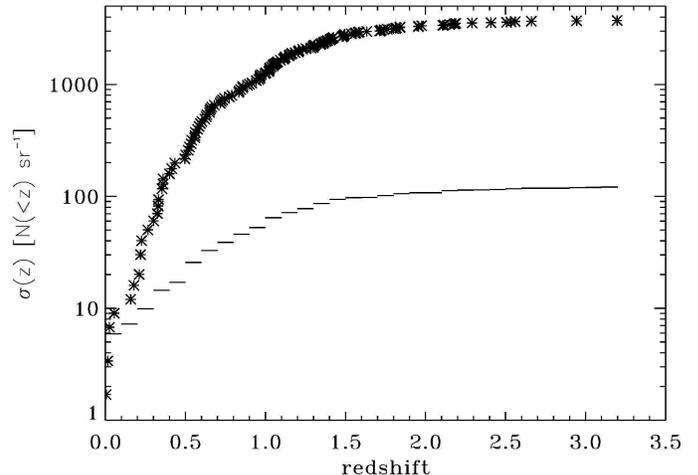}
\caption{The cumulative density function $\sigma(z)$ vs. redshift for FSRQ blazars.  The normalization of $\sigma(z)$ is determined as described in \S \ref{dev}.  A polynomial fit to $\sigma\!(z)$ is used to determine $\rho\!(z)$ by equation \ref{rhoeqn}.  We also show the cumulative number of observed FSRQs for redshift bins of size 0.1 for comparison.  It is seen that the raw data is significantly biased and the reconstructed intrinsic redshift distribution for FSRQs is very different than the observed one. }
\label{sigma}
\end{figure} 

\begin{figure}
\includegraphics[width=3.5in]{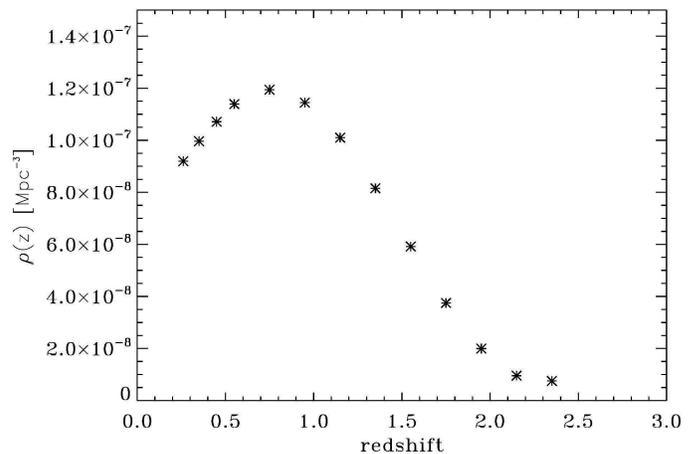}
\caption{The density evolution $\rho(z)$ vs. redshift for FSRQ blazars.  $\rho(z)$ is defined such that $\sigma(z')=\int_0^{z'} \rho(z) \, dV/dz \, dz$.  The normalization of $\rho(z)$ is determined as described in \S \ref{dev}, and polynomial fits of $\rho(z)$ to $z$ are given there.  At redshifts greater than 2.5, the smaller number of objects renders the determination of $\rho(z)$ with errors too large to be meaningfully plotted. }
\label{rholog}
\end{figure}

The normalization of $\rho(z)$ is determined by equation \ref{inteq2}, with the customary choice of $\int_{L_{\rm min}'}^{\infty} {\psi(L') \, dL'}=1$.  Figures \ref{sigma} and \ref{rholog} show the cumulative and differential density evolutions, respectively. The number density of FSRQ is seen to peak at around redshift 0.75.  This can be compared with the more complicated parameterization used in \cite{Marco} where the number density as a function of luminosity can be calculated.  For the median luminosity of objects in the sample (5.89$\times$10$^{47}$ erg sec$^{-1}$) the peak would be at $z$=0.8, while for the mean luminosity (2.47$\times$10$^{48}$ erg sec$^{-1}$) the peak is at $z$=1.7.  We see from Figure \ref{sigma} that the raw data is significantly biased and the reconstructed intrinsic redshift distribution for FSRQs is very different than the observed one.  The density evolution $\rho(z)$ can be fit quite well with a broken third-order polynomial in $z$, with coefficients shown in Table 1.

\section{Local Gamma-ray Luminosity Function} \label{ll}

One can use the local (redshift evolution taken out, or 'de-evolved') luminosity to determine the local cumulative  distribution $\Phi_{\rm L_{\gamma}'}\!(L_{\gamma}')$, where the prime indicates that the redshift evolutions have been taken out.  The cumulative distribution is related to the differential LF $\psi_{\rm L_{\gamma}'}\!(L_{\gamma}')$ by

\begin{figure}
\includegraphics[width=3.5in]{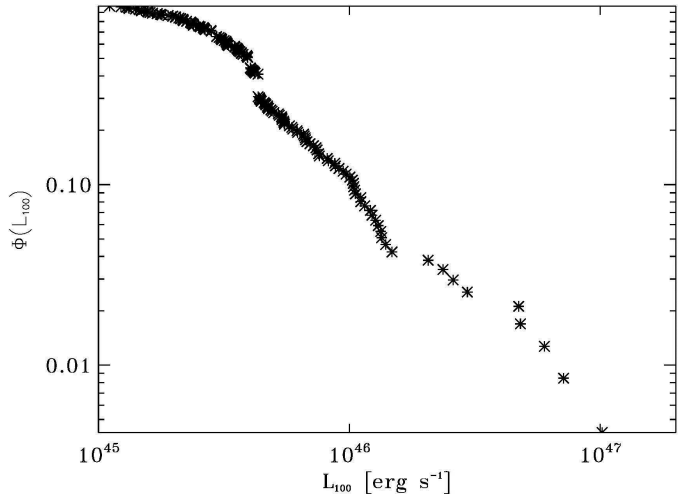}
\caption{The cumulative local ($z$=0) gamma-ray luminosity function $\Phi_{\rm L_{\gamma}'}\!(L_{\gamma}')$ for FSRQ blazars, as determined in \S \ref{ll}.   }
\label{phi}
\end{figure} 

\begin{figure}
\includegraphics[width=3.5in]{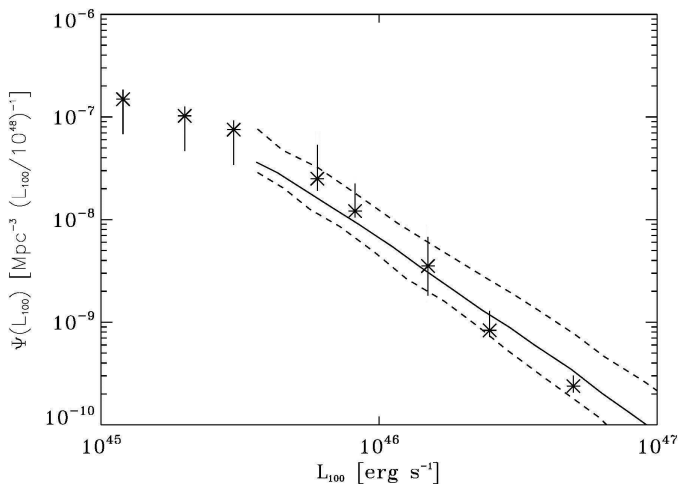}
\caption{The differential local ($z$=0) gamma-ray luminosity function $\psi_{\rm L_{\gamma}'}\!(L_{\gamma}')$ for FSRQ blazars, as determined in \S \ref{ll} (stars).  The error bars result from a propagation of the 1$\sigma$ variation in $k_{\rm L}$ and the uncertainty in the differentiation of $\Phi_{\rm L_{\gamma}'}\!(L_{\gamma}')$.  We overplot the $z$=0 FSRQ luminosity function as reported by \citet{Marco} (solid line) with 1$\sigma$ errors (dashed lines). }
\label{z0}
\end{figure}

\begin{figure}
\includegraphics[width=3.5in]{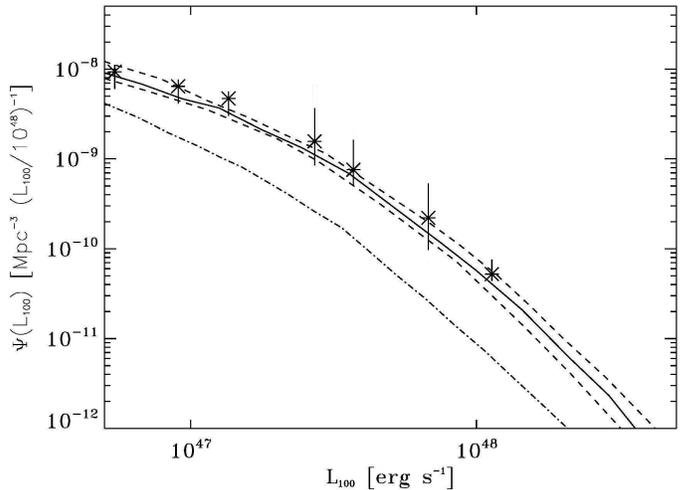}
\caption{The $z$=1 gamma-ray luminosity function $\Psi_{\rm L_{\gamma}}\!(L_{\gamma},z=1)$ for FSRQ blazars, as given by equation \ref{lumeq} (stars).  The error bars result from a propagation of the 1$\sigma$ variation in $k_{\rm L}$ and the uncertainty in the differentiation of $\Phi_{\rm L_{\gamma}'}\!(L_{\gamma}')$.  We overplot the $z$=1 FSRQ luminosity function as reported by \citet{Marco} (solid line) with 1$\sigma$ errors (dashed lines), and as reported by \citet{I10} (dash-dot line). }
\label{z1}
\end{figure} 

\begin{equation}
\Phi_{a'}\!(L_{a}') = \int_{L_{a}'}^{\infty} {\psi_a'\!(L_{a}'') \, dL_{a}''}
\end{equation}
Following \citet{P92}, \citet{QP1}, and \citet{QP2}, $\Phi_{a'}\!(L_{a}')$ can be calculated by

\begin{equation}
\Phi_{a'}\!(L_{a}') = \prod_{k}{\left(1 + {1 \over n(k)}\right)}
\label{phieq}
\end{equation}
where $k$ runs over all objects with a luminosity greater than or equal to $L_a$, and $n(k)$ is the number of objects with a luminosity higher than the luminosity of object $k$ which are in object $k$'s associated set, determined in the same manner as in \S \ref{evsec}. 

As before, the objects' luminosities, as well as the luminosity limits for inclusion in the associated set for given redshifts, are scaled by taking out factors of $g_{\rm L}\!(z)$, with $k_{\rm L}$ determined in \S \ref{evs}.  We use the notation $L \rightarrow L' \equiv L/g_{\rm L}(z)$ .  For the local distribution functions, we use the customary normalization involving the differential distribution $\int_{L_{\rm min}'}^{\infty} {\psi(L'') \, dL''}=1$.  This normalization may be biased by around 10\% due to  variability as discussed in BP1.  Figure \ref{phi} shows the local cumulative LF for FSRQ blazars.  

We can compute the differential local ($z$=0) gamma-ray LF $\psi_{\rm L_{\gamma}'}\!(L_{\gamma}')$, and combining it and the density evolution, we can compute the LF at any redshift $\Psi_{\rm L_{\gamma}}\!(L_{\gamma},z)$, and these can be compared to other determinations in the literature.  Figure \ref{z0} shows the differential local ($z$=0) gamma-ray LF along with that reported by \citet{Marco}.  It is seen that the determinations agree in the region of $L_{\gamma}'$ where they overlap.  We see that the results here access the $z$=0 LF for lower values of $L_{\gamma}'$ than reported in \citet{Marco}.  This follows from the rapid luminosity evolution seen here, where an object at redshift 1 will have its luminosity de-evolved by a factor of around 45 to redshift 0, uniform across luminosities.  In the luminosity-dependent density evolution scenario modeled in \citet{Marco} low luminosity sources are found to evolve slower and thus get de-evolved by a smaller factor. In principle different methods should find the same solution and this means sampling very similar de-evolved luminosities. In practice this rarely happens if the number of sources is small.  The results are somewhat complimentary in this manner, and show a flattening of the $z$=0 LF at lower luminosities.  

The LF at an arbitrary redshift $\Psi_{\rm L_{\gamma}}\!(L_{\gamma},z)$ can be computed by equation \ref{lumeq}.  In Figure \ref{z1} we plot the $z$=1 gamma-ray LF as determined here, along with those reported by \citet{Marco} and \citet{I10}.  It is seen that the determination of the LF here closely matches that in \citet{Marco} in spite of the different models considered (luminosity and density evolution versus luminosity dependent density evolution).  It is interesting to note that in both the $z$=1 LF and the integrated luminosity of FSRQs over the history of the Universe --- see \S \ref{bgnd} --- the results are consistent.

Since in \S \ref{evsec} we found that there is no redshift evolution in the photon index $\Gamma$ (i.e. $k_{\Gamma}$=0), the distribution we found for $\Gamma$ in BP1 averaged over redshifts is also valid at any individual redshift.  From BP1, the distribution of $\Gamma$ for FSRQs can be well described a Gaussian with a mean of 2.52$\pm$0.08 and a 1 $\sigma$ width of 0.17$\pm$0.02, a result in agreement with \citet{Marco0}.

\section{Contribution of FSRQs to the extragalactic gamma-ray background} \label{bgnd}

Given the luminosity and photon index evolutions as determined in \S \ref{evs}, the density evolution as determined in \S \ref{dev}, and the cumulative local LF as determined in \S \ref{ll}, we can calculate the integrated contribution of FSRQs of given redshifts and luminosities to the extragalactic gamma-ray background (EGB).  The contribution will be
\begin{eqnarray}
\mathcal{I}_{\rm \gamma : FSRQs}=\int_{z} \, dz \, \int_{L_{\gamma}} \, dL_{\gamma} \, \int_{\Gamma} \,  d\Gamma\, { {L_{\gamma}}  \over {4 \pi {D_L}^2 K_L(\Gamma,z) }  }
\\
\nonumber  \rho(z) \, { {dV} \over {dz}} \, { {\psi_{\rm L_{\gamma}} (L_{\gamma})} \over {g_{\rm L}(z)} }\, h(\Gamma) 
\end{eqnarray}
where $h(\Gamma)$ is the differential distribution of $\Gamma$ for FSRQs discussed in \S \ref{ll} and determined in BP1.  $\psi_{\rm L_{\gamma}}$ here is the luminosity part of the overall LF, including the redshift evolution of the luminosity (equivalent to $\psi_a\!(L_{a}/g_{a}\!(z) , \eta_a^j)$ in equation \ref{lumeq}).  We note that because $\Gamma$ is uncorrelated or only very weakly correlated with flux as determined in BP1, and that $h(\Gamma)$ is independent of reshift while $\psi_{\rm L_{\gamma}}(L_{\gamma})$ is strongly redshift dependent, $h(\Gamma)$ is likely, or can be well approximated by, an independent distribution.\footnote{This consideration also applies to equation \ref{inteq2} }.  We integrate $\psi_{\rm L_{\gamma}}(L_{\gamma}) dL_{\gamma}$ by parts, noting that
\begin{equation}
\nonumber \int_{L_{\gamma}} \psi_{\rm L_{\gamma}} (L_{\gamma}) \times L_{\gamma} \, dL_{\gamma} = \Phi_{\rm L_{\gamma}}(L_{\gamma}) \times L_{\gamma} \, \arrowvert_{L_{\gamma}} + \, \int_{L_{\gamma}} \,  \Phi_{\rm L_{\gamma}}(L_{\gamma})\, dL_{\gamma} 
\end{equation}
to obtain the dependence on the cumulative LF $\Phi_{\rm L_{\gamma}}(L_{\gamma})$
\begin{eqnarray}
\nonumber \mathcal{I}_{\rm \gamma : FSRQs} = \int_{z} \, dz \, \int_{\Gamma} \,  d\Gamma\, { {1}  \over {4 \pi {D_L}^2 K_L(\Gamma,z) g_{\rm L}(z) }  } 
\\
\nonumber  \rho(z) \, { {dV} \over {dz}} \, \left[\Phi_{\rm L_{\gamma}}(L_{\gamma}) \times L_{\gamma} \, \middle\arrowvert_{L_{\gamma}} + \, \int_{L_{\gamma}} \,  \Phi_{\rm L_{\gamma}}(L_{\gamma})\, dL_{\gamma} \right]  \, h(\Gamma) .
\end{eqnarray}
Stating this in terms of the local cumulative LF $\Phi_{\rm L_{\gamma}'}(  L_{\gamma}')$
\begin{eqnarray}
\mathcal{I}_{\rm \gamma : FSRQs} = \int_{z} \, dz \, \int_{\Gamma} \,  d\Gamma\,  { {1}  \over {4 \pi {D_L}^2 K_L(\Gamma,z) g_{\rm L}(z) }  }  \, \rho(z) \, { {dV} \over {dz}} \, \, \, \, \, \, \, \, \, 
\\
\nonumber  \left[\Phi_{\rm L_{\gamma}'}\left( {{ L_{\gamma} } \over {g_{\rm L}\!(z)} },z \right) \times L_{\gamma} \, \middle\arrowvert_{L_{\gamma}} + \int_{L_{\gamma}} \,  \Phi_{\rm L_{\gamma}'} \left( {{ L_{\gamma} } \over {g_{\rm L}\!(z)} },z \right)\, dL_{\gamma} \right]  \, h(\Gamma) 
\label{bgndcont}
\end{eqnarray}
Because we obtain the cumulative local LF $\Phi_{\rm L_{\gamma}'}$ and photon index distribution directly from the data, this technique allows us to calculate the cumulative total output of FSRQs directly, non-parametrically.  Specifically, we can approximate the integral directly by performing a Reimann sum with bins of luminosity and redshift, with the values of $\Phi_{\rm L_{\gamma}'}$ and $\rho(z)$ at any value of $L_{\gamma}$ and $z$ available via simple interpolation of the distribution functions that are the direct output of the methods.  It is simpler to use logarithmic bins of luminosity, in which case $dL_{\gamma} = L_{\gamma} d(\ln L_{\gamma})$.  

If one is integrating over FSRQs of all luminosities the surface term $\Phi_{\rm L_{\gamma}'}( {{ L_{\gamma} } \over {g_{\rm L}\!(z)} },z) \times L_{\gamma} \arrowvert_0^{\infty}$ is zero because $\Phi_{\rm L_{\gamma}'}(\infty)=0$ and one is left with
\begin{eqnarray}
\mathcal{I}_{\rm \gamma : FSRQs} = \int_{z} \, dz \, \int_{\Gamma=-\infty}^\infty \,   d\Gamma\, \int_{L_{\gamma}=0}^\infty \, dL_{\gamma} 
\\
\nonumber { {1}  \over {4 \pi {D_L}^2 K_L(\Gamma,z) g_{\rm L}(z) }  } \,  \rho(z) \, { {dV} \over {dz}} \, \Phi_{\rm L_{\gamma}'} \left( {{ L_{\gamma} } \over {g_{\rm L}\!(z)} },z \right) \, h(\Gamma) 
\label{bgndcont2}
\end{eqnarray}
We note that at as $z$ approaches large values $\rho(z) \rightarrow 0$, and also that as $L_{\gamma}$ approaches large values $\Phi_{\rm L_{\gamma}'} \rightarrow 0$, so the contributions from high very high redshifts and very large luminosities is vanishing.  Integrating over all redshifts and luminosities, we find that $\mathcal{I}_{\rm \gamma : FSRQs}$=7.6 (+4.0/-1.1) $\times$10$^{-4}$ MeV cm$^{-2}$ sec$^{-1}$ sr$^{-1}$.  The uncertainty results primarily from uncertainty in the value of $k_{\rm L}$, as well as uncertainty in the mean value of the photon index distribution for FSRQs.  

This value of this total $\mathcal{I}_{\rm \gamma : FSRQs}$ can be compared with the total energy density of the EGB, i.e. the total gamma-ray output of the Universe in the range from 100 MeV to 100 GeV, which was determined by {\it Fermi} in \citet{Fermibgnd}\footnote{Authors often divide this radiation into two parts, one consisting of the contribution of resolved sources and a second ``diffuse'' component, e.g. \citet{Fermibgnd}.  However, the most relevant comparison for the total emission from a class of object is with the total of these two, because which sources are declared to be resolved is a function of the properties of any particular survey.  Additionally, the most relevant quantity for answering astrophysical questions is the total contribution of a population relative to the total photon output of the Universe in that waveband. } to be $\mathcal{I}_{\gamma}$=4.72 (+0.63/-0.29)$\times$10$^{-3}$ MeV cm$^{-2}$ sec$^{-1}$ sr$^{-1}$, assuming a photon index for the background of $\Gamma$=2.4.

We see that FSRQ blazars in toto contribute 16 (+10/-4)\% of the total EGB, emphasizing that the EGB as defined here is the total gamma-ray output of the Universe in the range from 100 MeV to 100 GeV.  This is consistent with the result we obtained in BP1 where we estimated the contribution of {\it all} blazars (both FSRQs and BL Lacs) to be between 39\% and 100\% of the EGB if extrapolated to zero flux, and between 39\% and 66\% if extrapolated to a more reasonable lower limit flux of 10$^{-12}$ photons cm$^{-2}$ sec$^{-1}$.  The result here for FSRQs is also in agreement to within errors with the result of \citet{Marco} who find the contribution of all FSRQs to the total EGB to be 21.7 (+2.5/-1.7)\%.  The larger uncertainty presented here results from the uncertainty discussed above, as well as considering the full uncertainty range in $\mathcal{I}_\gamma$ reported by \citet{Fermibgnd}.

Some authors have suggested that blazars could be the primary source of the EGB \citep[e.g.][]{SV11,A10} and the results of BP1 did not rule this out, while the results presented in \citet{Marco}, \citet{Marco0}, and \citet{MH11} favor blazars being one of several important classes of sources.  Other possible source populations for the EGB include starforming galaxies, which have been proposed as a possible significant contributor to the EGB by e.g. \citet{SV11}, \citet{Fields10}, and \citet{L11}, although this is countered by \citet{M11}, radio galaxies \citep[e.g.][]{I11}, and non-blazar AGN \citep[e.g.][]{IT09,IT11}.

\section{Discussion} \label{disc}

We have used a rigorous method to calculate non-parametrically and directly from the data the redshift evolutions of the gamma-ray luminosity and photon index, as well as the density evolution, gamma-ray LF, and contribution to the EGB of FSRQ blazars.  We use a data set consisting of the FSRQs in the {\it Fermi}-LAT first year extragalactic source catalog with $TS \geq 50$ and which lie at Galactic latitude $\vert b \vert \geq 20^{\circ}$, with spectroscopically determined redshifts provided by \citet{Shaw}.  The method employed accounts robustly for the pronounced data truncation introduced by the selection biases inherent in the {\it Fermi}-LAT observational catalog.  The reliability of the methods employed has been demonstrated and discussed in \citet{QP1}, \citet{QP2}, and the appendix of BP1.   We note that since spectroscopic redshifts are available for virtually all of the FSRQs in the {\it Fermi}-LAT first year extragalactic source catalog, there is not a relevant limiting optical flux which must be considered, and the only relevant truncation in the data is that arising from gamma-ray flux and photon index.

In \S 3.3 of BP1 we discuss the sources of error that may affect these determinations, including measurement uncertainties, blazar variability, and source confusion.  As discussed there, in the determination of the contribution of FSRQs to the EGB, these will be sub-dominant to the uncertainty resulting from the range of mean values of the photon index distribution that we consider, as well as the high redshift end of the density evolution function.  Subtler issues affecting the derived distributions may arise because of the finite bandwidth of the {\it Fermi}-LAT and lack of complete knowledge of the objects' spectra over a large energy range and deviations from simple power laws.  This has the greatest potential to effect determinations of the photon index distribution, which we do not determine in this work but carry over from BP1.  However the bandwidth 100 MeV to 100 GeV is wide enough that the contribution of sources which peak outside of this range to the LF and evolution, density evolution, and the EGB in this energy range will be small. 

We find that FSRQs have a strong luminosity evolution with redshift, well characterized (at low redshifts) by the evolution factor (1+$z$)$^{k_L}$ with $k_L$=5.5$\pm$0.5.  This, along with positive evolutions in other wavebands such as optical and radio \citep[e.g.][]{QP2} and X-ray \citep[e.g.][]{Aird10}, favors models in which at higher redshifts AGN systems featured on average more massive black hole and accretion disk systems, and/or more rapidly rotating black holes.  We find that FSRQs do not exhibit appreciable photon index evolution with redshift, indicating that the mean spectrum of accelerated high energy particles from AGN central engines has remained constant over the history of the Universe.  Given these evolutions, the density evolution, and the local LF and photon index distribution, we determine that the total energy density from FSRQs at all redshifts and luminosities is $\mathcal{I}_{\rm \gamma : FSRQs}$=7.6 (+4.0/-1.1) $\times$10$^{-4}$ MeV cm$^{-2}$ sec$^{-1}$ sr$^{-1}$, which is 16 (+10/-4)\% of the total EGB.  This indicates that FSRQ blazars are a significant, but not dominant, component of the EGB.

\acknowledgments

The authors acknowledge support from NASA-Fermi Guest Investigator grant NNX10AG43G.  Additionally the authors thank Marco Ajello of the {\it Fermi}-LAT collaboration.

\appendix

\begin{figure}
\includegraphics[width=3.5in]{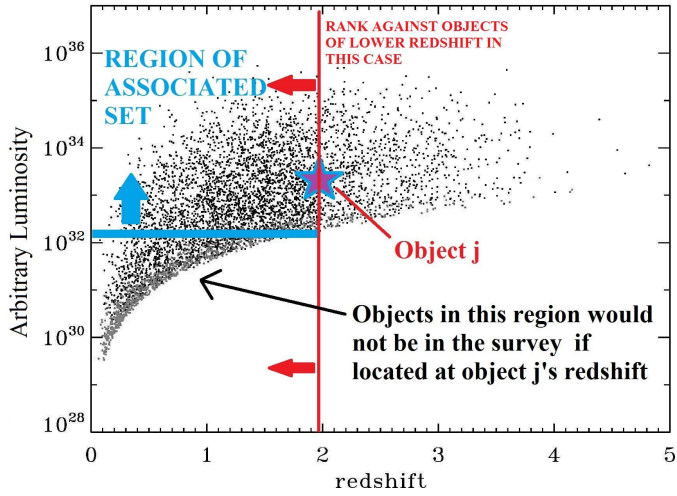}
\caption{A simple case to demonstrate the general concept of an associated set applied to truncated data.  A breif discussion is provided in the Appendix. }
\label{f10}
\end{figure}

In order to illustrate the general concept of an associated set applied to truncated data, in Figure \ref{f10} we show the simplest case of two-dimensional luminosity-redshift data truncated by a universal flux limit for the survey.  Suppose one were performing a statistical test to determine correlation among the variables, with luminosity as the dependent variable and redshift as the independent variable.  In order to form an unbiased set for comparison with object $j$ for this test, one should take only those objects that would be in the sample if they were at object $j$'s redshift.  With the Kendall Tau Test, one ranks objects in the dependent variable against those with either a higher or lower value of the independent variable.  In this case, the ranking would have to be against those with a lower value of redshift, in order for the associated sets themselves to be unbiased.  The associated sets for an object will be more complicated if there are additional variables, and/or if the truncation limit is not universal.  Both are the case in this work, where both luminosity and photon index are important, and where the truncation is a function in the flux-index plane.

\clearpage

\end{document}